\documentclass[10pt]{article}
\usepackage{amsmath}
\usepackage{graphicx}
\usepackage{amssymb}
\usepackage[utf8]{inputenc}
\usepackage{csquotes}
\usepackage{dcolumn}
\usepackage{bm}
\usepackage{orcidlink}
\usepackage{amsmath}
\usepackage{subcaption}
\usepackage[margin=0.88in]{geometry}
\usepackage{float}
\usepackage[sort&compress,numbers]{natbib}
\usepackage[T1]{fontenc} 
\usepackage{hyperref}
\hypersetup{
	colorlinks=true,
	linkcolor=blue,
	citecolor=red}

\title{Cosmology with non-linear barotropic Israel-Stewart fluid with causal relaxation time}

\author{Vishnu A Pai$^{1,*}$ \orcidlink{0000-0003-4161-3383} and Titus K Mathew$^{1,\ddagger}$  \orcidlink{0000-0002-9307-3772}\\ \small\textit{$^1$Department of Physics, Cochin University of Science and Technology, Kochi--682022, Kerala, India}\\ \small vishnuajithj@cusat.ac.in$^{*}$, titus@cusat.ac.in$^{\ddagger}$}

\date{\today}

\begin{document}
	\maketitle
\begin{abstract}
	We derive an extended expression for the relaxation time of a barotropic Israel–Stewart (IS) fluid using the non-linear causality constraint, and propose a new formulation for modeling causal viscous dissipation in barotropic fluids. With this generalized relaxation time, the non-linear IS equation simplifies to a first-order non-linear expression connecting bulk viscous pressure and energy density, which remains valid in any homogeneous and isotropic spacetime. In the case of spatially flat Friedmann universe, adopting this extended relation in the generalized non-linear IS theory, provides new class of analytical solutions in both, the linear, and the non-linear regimes. We also find that, the resulting effective equation of state in the linear regime naturally reproduces the generalized polytropic form which is often introduced phenomenologically in literature. Resulting dynamical implications are investigated and the constraints necessary for ensuring an acceptable evolutionary behavior for the fluid are determined. A detailed dynamical system analysis of the coupled Einstein–Israel–Stewart (EIS) system is also performed. Finally, we solve the coupled EIS equations numerically, and show that the model can support a transient Hubble slow-roll expansion phase with a smooth exit to a radiation-dominated universe, which is challenging to obtain in standard inflationary models.
\end{abstract}

\section{Introduction}

Relativistic dissipative hydrodynamics provides the theoretical foundation for describing and modeling non-equilibrium processes in relativistic fluids, encompassing applications from the early Universe to relativistic heavy-ion collisions and compact astrophysical objects \cite{Rocha:2023ilf,10.1093/mnras/stab1384,Romatschke:2017ejr}. In cosmology, it has been shown that, bulk viscous stresses can contribute to entropy generation \cite{Maartens1995}, particle creation \cite{Zimdahl1996}, and the damping of anisotropies in the early universe \cite{Barrow1986,Lima1996}. Moreover, it was also shown that, under suitable conditions, an interacting multi-component medium can be recast as an effective bulk viscous fluid \cite{10.1093/mnras/280.4.1239,10.1093/mnras/199.4.1137}. These profound cosmological implications of bulk viscosity have attracted considerable attention, particularly its ability to drive acceleration expansion of the Universe \cite{Giovannini1999,Velten2012,Cruz2015,Sasidharan2015}. 

The theoretical formulation of relativistic dissipative fluid dynamics has undergone substantial development over the past decades. First-order theories proposed by Eckart \cite{Eckart1940}, and later by Landau \& Lifshitz \cite{Landau1987}, were first-order relations connecting thermodynamic fluxes and gradients of macroscopic variables. However, in spite of being direct relativistic extension of the standard Navier-Stokes relation, these formulations were shown to violate causality \cite{IsraelStewart1979} and predict solutions with unstable final equilibrium states \cite{PhysRevD.31.725}. To overcome these drawbacks, Israel and Stewart generalized the above approach by considering second-order corrections with additional relaxation type degree of freedoms in the entropy flux equation \cite{Israel1976,IsraelStewart1979}. The resulting formulation, called the Israel-Stewart theory (IS), forms a thermodynamically consistent framework with relaxation-type equations for viscous pressure, which thereby ensures causality within the near-equilibrium limit . However, the IS theory did not provide a general proof of causality and stability for fluids, that are driven far from equilibrium, or when nonlinear dissipative effects are dominant. This long-standing problem was recently solved in \cite{PhysRevLett.122.221602}, where the authors derived the necessary and sufficient conditions under which causality, stability, and hyperbolicity are satisfied in general non-linear IS theories that are dynamically coupled to Einstein equations. This establishes the mathematical consistency of relativistic hydrodynamics beyond the linear regime, and provides a robust foundation for studying strongly non-equilibrium phenomena in relativistic hydrodynamic systems.

Despite its wide ranging applications, in most cases, IS framework in its standard formulation requires phenomenological prescriptions for the relaxation time ($\tau$) and the viscous coefficient ($\zeta$), particularly in cosmological settings. In most cases, the dissipative variables are considered to be barotropic, and are hence modeled as a function of the fluid’s energy density \cite{Weinberg,PhysRevD.8.4231,udey}. While these basic assumptions greatly simplify analytical and numerical calculations, it generally does not ensure that the resulting theory respects causality. The first explicit causal relation for the relaxation time was derived by Maartens in  \cite{Maartens1996}, which established a direct connection between dissipative variables within the near-equilibrium regime. However, since this relation is obtained under the assumption of small deviations from equilibrium, adopting the same relation in far-from-equilibrium regimes may not ensure causality. In cosmological contexts, where rapid expansion and significant departures from local equilibrium may occur, these simplifications can affect the accurate modeling of bulk dissipative processes \cite{Kremer2003,PeraltaRamos2012,Barrow1990}. Furthermore, it is to be noted that, viscous driven accelerated expansion of the universe is an inherently far-from-equilibrium behavior. This begs the question: How will the general causality constrain derived in \cite{PhysRevLett.122.221602}, affect the relaxation dynamics of relativistic dissipative fluids beyond linear regime ? 

In this work, we derive an extended expression for the relaxation time of a barotropic, nonlinear IS fluid directly from the general causality condition proposed in \cite{PhysRevLett.122.221602}, such that the resulting formulation guarantees causality under arbitrary departures from equilibrium. We will show that, for a barotropic fluid, adopting the aforementioned expression for relaxation-time can reduce the nonlinear IS equation to a first-order non-linear differential equation connecting the bulk viscous pressure and the energy density, which remains valid in any homogeneous and isotropic space-times. This reduction renders the theory analytically tractable while also preserving the necessary causal structure. We will then apply this IS equation in the context of spatially flat FLRW space-time, and show that this formulation yields new class of exact analytical solutions that resembles the generalized polytropic equations of states, which are often introduced phenomenologically \cite{KAMENSHCHIK2001265,doi:10.1142/S0218271809015795,PhysRevD.71.063004,Chavanis:2012uq,DUNSBY2024101563,Chavanis_2018}. Finally, we will also show that, the resulting dynamical equation can also enable a transient viscous driven early-inflationary phase that naturally evolve into radiation-dominated epoch and remains observationally consistent.


\section{Non-linear Israel-Stewart theory with causal relaxation time}\label{2}
 
In homogeneous and isotropic space-times, one defines the energy momentum tensor and particle flux density of dissipative fluids with energy density `$\rho$', non-equilibrium pressure `$\mathcal{P}$', particle number density `$\mathfrak{n}$', as,
\begin{equation}\label{tmn}
	T^{\mu \nu}=\left(\rho+\mathcal{P}\right)u^{\mu}u^{\nu} +\mathcal{P}g^{\mu \nu} \quad\;\; \textbf{:}\;\;\quad \mathfrak{n}^{\mu}=\mathfrak{n} u^{\mu} 
\end{equation}
Here, $\mathcal{P}=p+\Pi$, with $p$ as the equilibrium pressure, and $\Pi$ as the bulk viscous pressure, $u^{\mu}$ is the particle four-velocity in the instantaneous comoving frame, and $g^{\mu \nu}$ is some arbitrary space-time metric. In the context of general relativity, the equations of motion that govern the dynamics of such a fluids are \cite{PhysRevLett.122.221602},
\begin{align}
	u^{\mu}\nabla_{\mu}	\rho\;\;&=-\left(\rho+p+\Pi\right) \nabla_{\mu}u^{\mu}\label{cons}\\
	u^{\mu}\nabla_{\mu}	u_{\nu }\,&=-\frac{\left[\,\omega_\rho \nabla_{\nu}^{\mu}\Delta_{\mu}\rho + \omega_\mathfrak{n} \nabla_{\nu}^{\mu}\Delta_{\mu}\mathfrak{n} + \nabla_{\nu}^{\mu}\Delta_{\mu} \Pi\,\right]}{\rho+p+\Pi} \label{momen}\\
	u^{\mu}\nabla_{\mu}	\mathfrak{n} \;\;&=-\mathfrak{n}\,\nabla_{\mu}u^{\mu}\label{num}\\
	u^{\mu}\nabla_{\mu}	\Pi\;\, &=-\frac{1}{\tau_{\Pi}} \left[\, \Pi +\lambda \Pi^2+\zeta_{\Pi} \nabla_{\mu}u^{\mu}\,\right]\label{diss}\; 
\end{align}
Here, $\left(\omega_\rho, \omega_\mathfrak{n},  \zeta_{\Pi}, \tau_{\Pi},\lambda\right)$ are free parameters of the model that depend on the nature of the fluid, and in general, they are functions of the independent variables ($\rho, \mathfrak{n}$). In particular, $\omega_\rho=(\partial p/\partial \rho)_{\mathfrak{n}}$ is the squared adiabatic speed of sound in the fluid medium, $\omega_n=(\partial p/\partial \mathfrak{n})_{\rho}$ quantifies the  dependence of pressure on particle number density, `$\tau_\Pi$' represents the relaxation time of the fluid, `$\zeta_{\Pi}$' is the coefficient of bulk viscosity and `$\lambda$' corresponds to the coupling strength of the non-linear viscous pressure term, which we treat as a constant free parameter in the model\footnote{Note that $\lambda$ has the dimension of inverse energy density.}. The transport equation (\ref{diss}) constitutes a non-linear Israel–Stewart relation, originally proposed in \cite{PhysRevLett.122.221602}, that captures the essential physical characteristics of the system while retaining a concise form. Notably, in the linear regime ($\lambda=0$) it reduces to the standard Maxwell-Cattaneo relation, and in addition if the limit $\tau_{\Pi} \to 0$ is also satisfied, it approaches the relativistic Navier-Stokes relation. To close the system, one must then express the dissipative variables $\left(\omega_\rho, \omega_\mathfrak{n},  \zeta_{\Pi}, \tau_{\Pi}\right)$ in terms of the independent variables. Ideally, these relations should be derived from underlying microscopic interactions in the fluid. However, in most cases such a description is not readily obtained, and consequently one postulate these relations phenomenologically. 

In the general case of a barotropic bulk viscous fluid both, the equilibrium pressure, and the viscous pressure, depend only on the energy density of the fluid (and not on particle number density), and hence one often adopts the relations; $\Pi=\Pi(\rho)$, and $p=w_0\rho$. Here, $w_0$ is considered to be a constant free parameter\footnote{The parameter `$w_0$' is related to adiabatic index `$\gamma$' via the relation, $\gamma=1+w_0$.}, though in general, it can also vary with energy density. It is then easy to see that, owing to this choice of pressure we get, $\omega_\rho=w_0$ and $\omega_\mathfrak{n}=0$. However, obtaining the form of  $\zeta_{\Pi}$ and $\tau_{\Pi}$ is not that straightforward. In the linear regime, authors most often consider two different approaches in choosing $\zeta$ and $\tau$; One either treats them as independent barotropic variables \cite{Chimento:1997vy,PhysRevD.63.023501}, or alternatively, one relates them via the causality constraint in the linear regime \cite{Maartens1996}, and subsequently, postulates a barotropic bulk viscous coefficient \cite{PhysRevD.100.083524}. Regardless, in the non-linear regime, or in the cases where when $\tau$ and $\zeta_{\Pi}$ are allowed to depend also on $\Pi$,  both these cases may fail to satisfy causality. This is because, the linear constraint may not applicable in the non-linear regime. Hence, in such cases one must replace the linear causality constraint with the non-linear version derived in \cite{PhysRevLett.122.221602}. In that manuscript, authors proved that the set of equations (\ref{cons})-(\ref{diss}), coupled to the Einstein's field equation, represents a first-order symmetric hyperbolic system (FOSH) which is causal, stable and hyperbolic in the non-linear regime, provided the constraint given below is satisfied;
\begin{equation}
	\left[\frac{\zeta_{\Pi}}{\tau_{\Pi}} + \mathfrak{n}\,\omega_\mathfrak{n}\right]\frac{1}{\rho+p+\Pi}\,\leq\, 1-\omega_\rho
\end{equation}
In the present case, since $\omega_\mathfrak{n}=0$ and $\omega=w_0$, we can relate the effective relaxation time to viscous coefficient via the constraint relation (as in linear regime \cite{Maartens1996}) as;
\begin{equation}\label{tau}
	\tau_{\Pi}=\frac{\zeta_{\Pi}}{\epsilon_0\left(1-w_0\right)\left(\rho+p+\Pi\right)}\quad \textbf{:}\quad 0<\epsilon_0\leq1
\end{equation}
Notably, since the above definition of the relaxation time is directly derived from the causality constraint, the transport equation for bulk viscosity will remain causal at all times, even when fluid is far from equilibrium, i.e. $|\Pi|\approxeq(\rho+p)$. Consequently, we will refer this version of relaxation time as the `causal relaxation time' ($\tau_{c}$) and the associate fluid as BISF, i.e. Barotropic Israel Stewart Fluid having `$\tau_c$' relaxation time. Interestingly, this equation exactly reduces to standard relation in the linear regime when the fluid is in a near-equilibrium state, i.e. when $|\Pi|/(\rho+p)\ll1$, and can hence be considered as an extended version of the linear relation. Consequences of adopting this relaxation time can be inferred by analyzing Eqn. (\ref{diss}) subject to Eqn. (\ref{tau}). 
\paragraph*{\textbf{Far from equilibrium}:} In this regime, $|\Pi| \to \rho+p$ and as a result, $\tau_{c}$ increases significantly and may even diverge when $|\Pi| = \rho+p$. To bypass this singularity, one must propose suitable initial condition for the fluid, such that $|\Pi|$ always remains less than $\rho+p$, as we will show in the upcoming section. This implies three things; 
	\begin{itemize}
		\item For a finite value of $\Pi$, the rate of change of $\Pi$ at large relaxation times is small, and as a result, the fluid exhibits pseudo-plasticity\footnote{A pseudoplastic fluid is a material whose coefficient of bulk viscosity, changes as a function of the deformation rate \cite{PhysRevD.109.096040}}.
		\item Viscous driven de-Sitter expansion is a theoretical impossibility since $\rho+p+\Pi=0$ is non-achievable. However, a quintessence solution mimicking a quasi de-Sitter phase is plausible when, $\Pi\approx-\left(\rho+p\right)$.
		\item Phantom viscous fluids are also theoretically non-feasible. This is because, in the present case, the fluid must satisfy the constraints, $\rho+p+\Pi\geq0$, $w_0\geq0$ and $\zeta_{\Pi}\geq0$, at all times to ensure causality and uniqueness of solution \cite{PhysRevLett.122.221602}.
	\end{itemize}
\paragraph*{\textbf{Near-equilibrium:}} In the near-equilibrium state of the viscous fluid, we have $|\Pi| \ll \rho+p$. Consequently, $\tau_{c}$ has a small, yet finite value in this regime. From equation (\ref{diss}), this means that the contribution from both, the bulk viscous pressure, and its rate of change can be significant. From rheological point of view this corresponds to viscoelastic behavior \footnote{A viscoelastic fluid is a material whose stress exhibits a delay in the response to time-dependent deformation rates \cite{PhysRevD.109.096040}}.

Combining equations (\ref{cons}), (\ref{diss}) and (\ref{tau}) for a barotropic fluid, one obtains the non-linear ordinary differential equation (ODE),
\begin{equation}\label{ode}
	\frac{d\hspace{0.025cm}\Pi}{d \rho}=\epsilon_0\left(1-w_0\right)\left[1+\frac{\Pi\left(1+\lambda \Pi\right)}{\zeta_{\Pi}\; \nabla_{\mu}u^{\mu}}\right]
\end{equation}
This ODE is quite general as it is independent of the choice of space-time metric, and was obtained as a direct consequence of the two assumptions that; (a) the bulk viscous pressure is barotropic (b) the relaxation time is given by Eqn. (\ref{tau}). In the upcoming sections, we will analyze the behavior of this equation in the cosmological context, considering both linear and non-linear effects, and perform dynamical stability analysis of the coupled Einstein-Israel-Stewart system to find viable solutions.

\section{Spatially flat Friedmann universe with BISF}\label{3}

We will now investigate the dynamics of a spatially flat, homogeneous and isotropic universe, when the ideal fluid is replaced by the BISF modeled in the previous section. By adopting the standard Freedman-Lema$\hat{\text{i}}$tre-Robertson-Walkermetric,
\begin{equation}\label{gmn}
	ds^2=-c^2 dt^2+a(t)^2\left[\,dx^2+dy^2+dz^2\,\right],
\end{equation}
in the Einstein's field equations, and using Eqns. (\ref{tmn}), (\ref{gmn}), we get the Friedmann equations,
\begin{align}
	3H^2&=\rho\\
	2\dot{H}+3H^2&=-(p+\Pi)
\end{align}
Since we are considering a barotropic viscous pressure, the dissipative coefficient $\zeta_{\Pi}$ must also be barotropic, i.e. $\zeta_{\Pi}\sim\zeta_{\Pi}(\rho)$. The simplest, most straight forward ansatz that one could consider is then, $\zeta_{\Pi}(\rho)=\zeta_{0}\sqrt{\rho^{2n+1}}$, which is a popular form often considered in literature \cite{Maartens1996}. Adopting this relation in Eqn. (\ref{ode}), one can simplify the resulting ODE and obtain,
\begin{equation}\label{ode2}
	\frac{d\hspace{0.025cm}\Pi}{d \rho}=\epsilon+\eta\left[\frac{\,\Pi\left(1+\lambda \Pi\right)}{ \rho^{\,n+1}}\right].
\end{equation}
Here, we have defined $\epsilon=\epsilon_0\left(1-w_0\right)$, $\eta=\sqrt{3}\,\epsilon/\zeta_{0}$, and $\epsilon_0$ and $\zeta_0$, are constant real numbers\footnote{Note that, since $n$ is an unspecified constant real number, one has the freedom to appropriately choose the power of $\rho$ in the ansatz for viscous coefficient for better representation. Here we adopted $\sqrt{\rho^{\,2n+1}}$ so that final expression looks less tedious. Since $n$ is not specified, one might as well consider $\rho^m$, where $m=n+1/2$.}.

To solve this non-linear ODE, one needs to impose a suitable initial condition, and in cosmological context, the natural intuition is that the fluid starts off from an out-of-equilibrium state, and later approaches a stable equilibrium as it evolves in time. Therefore, it is expected that the end phase solution is a stable equilibrium state where the bulk dissipative fluxes necessarily vanish. In this article we will adhere to this intuitive picture. Hence, we consider the fluid to be evolving from an initial far-from-equilibrium state with $\rho=\Lambda_e$, and $\Pi_e=-\mathfrak{w}\Lambda_e$, towards a late-time stable state as energy density dilutes in time, i.e. acceptable solution must have $\Pi/\rho \to 0$ as $a \to \infty$ and $\rho \to 0$. Here, $0<\mathfrak{w}<1+w_0$, is a constant parameter that quantifies how far the fluid is from its equilibrium state when its evolution began. The upper bound for this parameter is obtained from the dominant energy condition, $\rho+p+\Pi\geq0$.
\begin{figure}
	\centering
	\includegraphics[width=0.5\columnwidth]{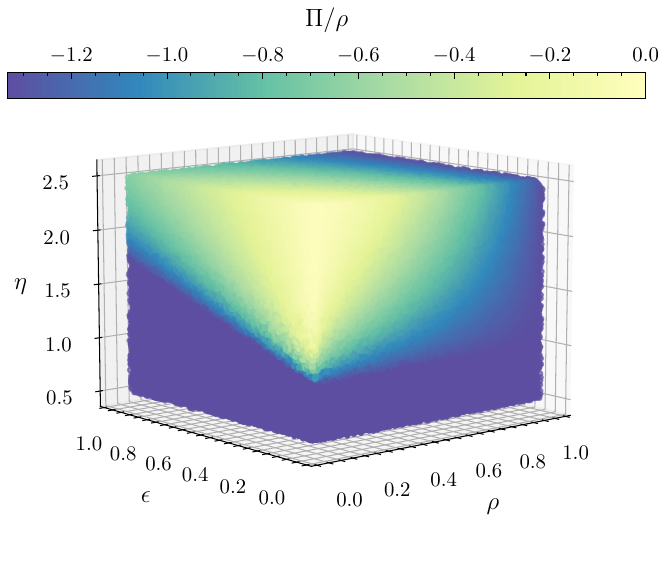}
	\caption{This is a 4D scatter plot depicting the evolution of $\Pi/\rho$ with three different parameters, corresponding to $n=0$ case of BISF in the linear regime.}
	\label{4DPL}
\end{figure}

\subsection{BISF in the linear regime}
In the linear regime, $\lambda=0$, and the resulting ODE can be exactly solved to obtain the following analytical solutions for cases (a) $n=0$, and (b) $n \neq 0$,
\paragraph{\textbf{(a) For $n=0$:}} In this case we get,
\begin{equation}\label{Pilinearn0}
	\Pi =\left[\frac{\epsilon}{1-\eta}\right]\rho+\mathbb{C}\; \rho ^{\;\eta}
\end{equation}		
Interestingly, the above equation connecting the viscous pressure with energy density is identical to the equation of state of a generalized polytroic fluid \cite{KAMENSHCHIK2001265,doi:10.1142/S0218271809015795,PhysRevD.71.063004,Chavanis:2012uq,DUNSBY2024101563,Chavanis_2018}. Applying the initial condition $\Pi_e=-\mathfrak{w}\Lambda_e$ at $\rho=\Lambda_e$, we get the exact solution as,
\begin{equation}
	\frac{\Pi}{\rho}=\frac{\epsilon}{1-\eta}-\left[\mathfrak{w}+\frac{\epsilon}{1-\eta}\right]\; \left[\frac{\rho}{\Lambda_e} \right]^{\;\eta-1} \quad \textbf{;}\quad \eta \neq 1
\end{equation}

By analyzing this equation, significant inferences can be drawn regarding the possible behaviors that the fluid could show, and assess which among them are physically viable ones. As mentioned earlier, the acceptable solutions are those in which the fluid is seen evolving towards the equilibrium state as the energy density dilutes. This means, as $\rho$ evolves from $\Lambda_e$ to zero, $|\Pi|/\rho$ must decrease from its initial value, and therefore, $\Pi/\rho$ must tend to zero as $\rho\to 0$. To visualize this clearly, we make a 4D plot of the above equation by considering $\epsilon$, $\eta$ and $\rho$ as the three axes of a coordinate system, and plotting $\Pi/\rho$ as a color gradient depicted via the colorbar, as seen in Fig. (\ref{4DPL}). In the plot, we have fixed\footnote{We have considered the case where fluid is initialized near to its maximum possible out-of-equilibrium state $\mathfrak{w}\approx1+w_0$. Recall that maximum value of $w_0$ is $1/3$. However, while comparing with actual observations, $\mathfrak{w}$ should be treated as a free parameter.} $\Lambda_e=1$ and $\mathfrak{w}=1.33$. Based on the colormap that we have used, the acceptability criterion stated above implies the following; \textit{As energy density ($\rho$) evolves from one to zero, the viable solutions are those for which the color gradates from dark-blue to bright-yellow}. Accordingly, from Fig. (\ref{4DPL}) we see that for having viable solutions, $\eta$ must be sufficiently large, while $\epsilon$ should be necessarily small. As the value of $\epsilon$ increases and approaches one, value of $\eta$  required for obtaining an acceptable solution becomes larger. At the same time, we also see that if value of $\eta$ becomes smaller than one, no value of $\epsilon$ (how much small it may be), will yield a solution that is evolving towards equilibrium. Furthermore, if we consider $\eta<1$, then from Eqn. (\ref{Pilinearn0}), we will obtain the scenario where, $\Pi/\rho\to-\infty$ as $\rho \to 0$, which is unacceptable. Hence in this case, for having a physically viable evolution for the bulk viscous fluid, it is mandatory that, $\eta>1$ and $\epsilon\ll1$. Also, in this domain of $\eta$ and $\epsilon$ values, one can show that;
\begin{equation}
	\lim_{\rho \to 0} \frac{\Pi}{\rho} \;\;\bigg|_{\eta>1}=\frac{\epsilon}{1-\eta}
\end{equation}
Hence, a small `$\epsilon$' is also needed to ensure that the viscous fluid is close to equilibrium state in the late phase.

\begin{figure}
	\centering
	\begin{subfigure}{0.495\columnwidth}
		\centering
		\includegraphics[width=0.98\columnwidth]{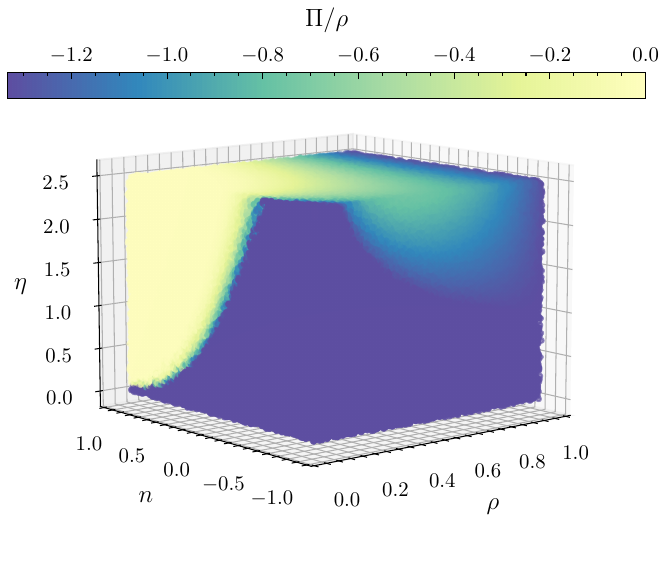}
		\caption{For different $n$ and $\epsilon=0.01$.}
		\label{4DPLn(fep)}
	\end{subfigure}
	\begin{subfigure}{0.495\columnwidth}
		\centering
		\includegraphics[width=0.98\columnwidth]{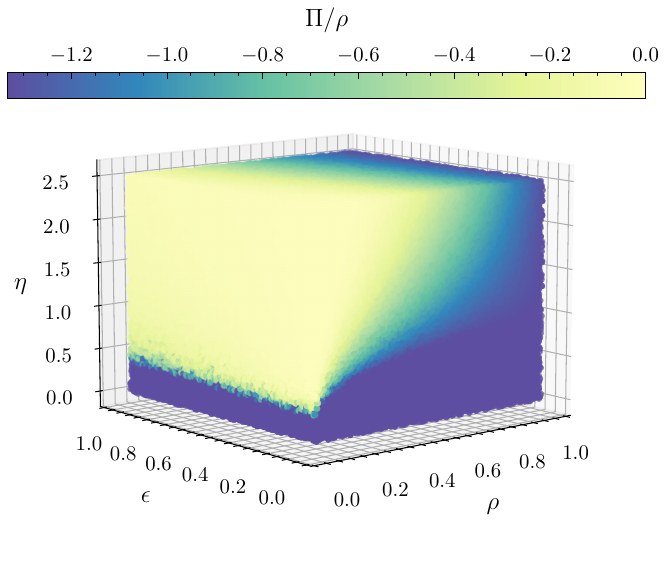}
		\caption{For different $\epsilon$ and $n=0.25$.}
		\label{4DPLn(fn)}
	\end{subfigure}
	\caption{These are 4D scatter plots showing the evolution of $\Pi/\rho$ with three different parameters, for $n\neq0$ case of BISF in the linear regime.}\label{4DPLn}
\end{figure}
\paragraph{\textbf{(b) For $n\neq0$:}} Surprisingly, the analytical solution to Eqn. (\ref{ode2}) can be obtain for any arbitrary value of $n$ except zero (in which case one uses the solution obtained in the previous case). This is derived via the method of integrating factors, which yields;
	\begin{align}
		\frac{\Pi}{\rho}&= \exp{\left[\frac{-\,\eta}{n\rho ^{\,n}}\right]} \left\{\mathbb{E}-\mathfrak{w}\left[\frac{\Lambda_e}{\rho}\right]\exp\left[\frac{\eta}{n\Lambda_e^{\;n}}\right]\right\}\\
		\mathbb{E}&=\frac{\epsilon}{\rho}\int_{\Lambda_e}^{\,\rho} \exp\left[\frac{\eta}{n \tilde{\rho}^{\,n}}\right]d\tilde{\rho}\notag
	\end{align}
Here we have used the same initial condition, as in the previous case, and have integrated within the limit $\left(\rho,\Lambda_e\right)$. Exact behavior of this solution can be studied with the help of the plots in Fig. (\ref{4DPLn}). To plot figure (\ref{4DPLn(fep)}) we have taken the values, $\Lambda_e=1$, $\mathfrak{w}=1.33$, $\epsilon=0.01$, and allowed `$(n,\eta,\rho)$' to vary. Where as, figure (\ref{4DPLn(fn)}) is constructed by fixing $\Lambda_e=1$,  $\mathfrak{w}=1.33$, $n=0.25$ and varying `$(\epsilon,\eta,\rho)$'.

Analyzing these figures following inferences can be drawn;\\
\textbf{(i)} For $n<0$, the viscous fluid initially evolves towards the equilibrium state (as indicated by a change in color from dark-blue to light-yellow in Fig. (\ref{4DPLn(fep)})) but later, as $\rho$ tends to zero, it is driven away from equilibrium state (as indicated by a color shift from light-yellow to dark-blue). This fact is true irrespective of the values of $\eta$ and $\epsilon$. Consequently, in this case, the dissipative fluid totally fails to approach equilibrium state in the late phase. \\
\textbf{(ii)} For $n>0$, fluid can evolve towards equilibrium state, provided $\epsilon\ll1$ and `$\eta$' is above a particular threshold which in-turn depends on the value of `$n$'. From Fig. (\ref{4DPLn(fep)}) it is clear that for $\epsilon\ll1$; large positive values of `$n$' can ensure viable solutions even for small $\eta$ values. Hence, $\eta>1$ is not a strict requirement to predict viable solutions in the cases where $n>0$.\\
\textbf{(iii)} As $n\to0$, the plot begins to align closely with the previous case in which $n=0$, i.e. Fig. (\ref{4DPL}). As expected, from Fig. (\ref{4DPLn(fn)}), we see that the inferences drawn in the earlier case will apply here also, i.e. viable solutions are obtained when, $\eta>1$ and $\epsilon\ll1$.
\begin{figure}
	\centering
	\begin{subfigure}{0.495\columnwidth}
		\centering
		\includegraphics[width=\columnwidth]{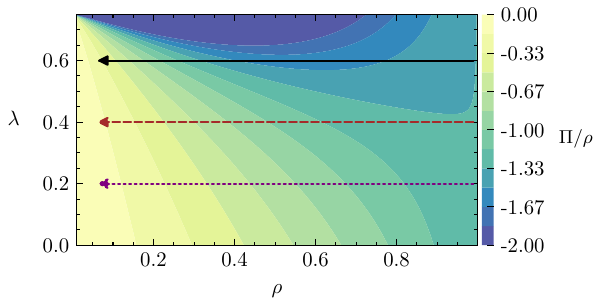}
		\caption{For $\eta=1.90$.}
		\label{contoura}
	\end{subfigure}
	\begin{subfigure}{0.495\columnwidth}
		\centering
		\includegraphics[width=\columnwidth]{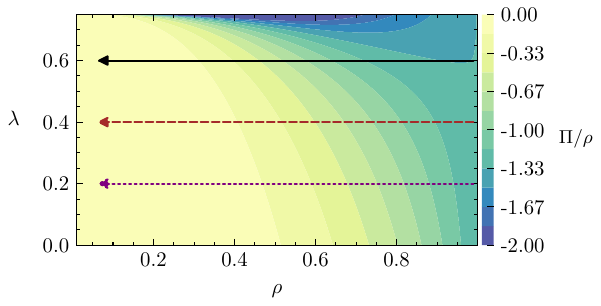}
		\caption{For $\eta=2.90$.}
		\label{contourb}
	\end{subfigure}
	\caption{These are contour plots showing the evolution of $\Pi/\rho$ with three different parameters, for $n\neq0$ case.}
\end{figure}

\subsection{BISF in the non-linear regime} 

To investigate the influence of non-linearity, we will restrict our analysis to the case where $\lambda \neq 0$ and $n = 0$ in Eqn. (\ref{ode2}). This reduces the number of free parameters in the model and simplifies the equations. Furthermore, this particular choice allows one to derive analytical solution for the bulk viscous pressure presented below,
\begin{align}
	\frac{\Pi}{\rho}&=-\frac{2\,\epsilon }{\mathcal{A}}\left\{\frac{\Gamma\left[\eta\right] J_{\eta -1}\left[\mathcal{A}\right]+\mathbb{C}\,\Gamma\left[-\eta\right]J_{-\left(\eta-1 \right)}\left[\mathcal{A}\right]}{\Gamma\left[\eta\right] J_{\eta }\left[\mathcal{A}\right]-\mathbb{C}\,\Gamma\left[-\eta\right] J_{-\eta }\left[\mathcal{A}\right]}\right\}\\
	\mathbb{C}&=\frac{\Gamma\left[\eta\right] }{\Gamma\left[-\eta\right] }\left\{\frac{\mathcal{B} \mathfrak{w} \,J_{\eta }\left[\mathcal{B}\right]-2 \epsilon \, J_{\eta -1}\left[\mathcal{B}\right]}{\mathcal{B} \mathfrak{w}\, J_{-\eta }\left[\mathcal{B}\right]+2 \epsilon \,J_{-\left(\eta-1\right) }\left[\mathcal{B}\right]}\right\}
\end{align}
where we have defined,
\begin{align}
	\mathcal{A}&=2\sqrt{\epsilon\eta\lambda\rho} \quad\quad\textbf{\&}\quad\quad \mathcal{B}=2\sqrt{\epsilon\eta\lambda\Lambda_e}\\
	\Gamma[z]&=\int_{0}^{\infty}x^{z-1}e^{-x} \,dx\quad\left(\textbf{Gamma Function}\right)
\end{align}
and, $J_{x}\left[y\right]$ represents the Bessel function of the first kind. Analysis of this equation leads to following inferences;\\
\textbf{(i)} Presence of the Gamma functions $\Gamma \left[\eta\right]$ and $\Gamma \left[-\eta\right]$ imposes a restriction on the parameter $\eta$, as the gamma function is undefined for negative integers. This is not that problematic since $\eta$ represents a combination of $\epsilon_0$, $w$, and $\zeta_{0}$, rather than a single parameter, and it is highly improbable that $\eta$ assumes integer values. \\
\textbf{(ii)} To ensure that $\mathcal{A}$, $\mathcal{B}$, and consequently $\Pi/\rho$, remains real valued, it is necessary that $\lambda>0$.

\begin{figure}
	\centering
	\includegraphics[width=0.5\columnwidth]{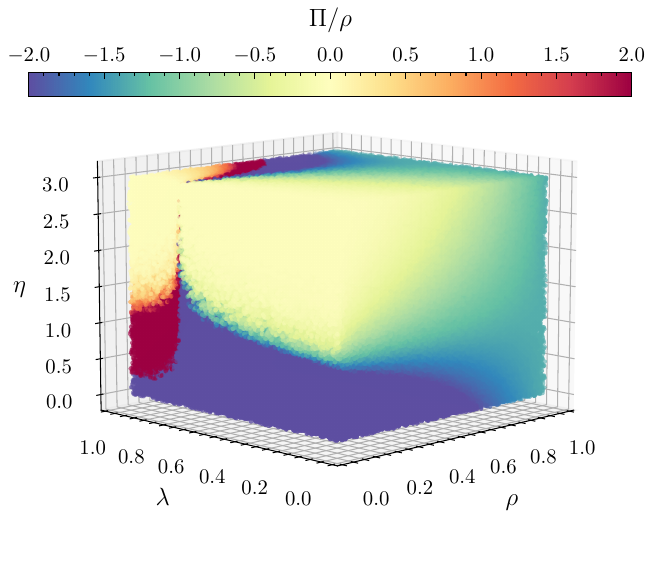}
	\caption{Scatter plot depicting the evolution of $\Pi/\rho$ with 3 model parameters, for linear, $n=0$ case of BISF.}
	\label{4DNL}
\end{figure}

\begin{figure}
	\centering
	\includegraphics[width=0.5\columnwidth]{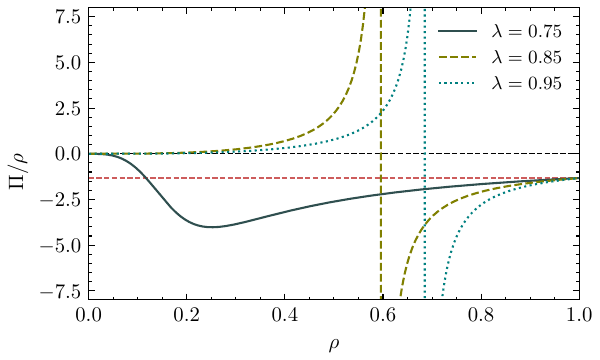}
	\caption{Plot depicting the evolution of $\Pi/\rho$ with energy density, for $\Lambda_e=1, \epsilon=0.01, \eta=4.15$ and three different $\lambda$ values. Dotted dashed red line is for $\Pi/\rho=-4/3$.}
	\label{PiNL}
\end{figure}
As done in the previous cases, additional constraints on the model parameters can be deduced by analyzing the conditions for viable solutions. This is done by referring to the plots (\ref{contoura}), (\ref{contourb}), and (\ref{4DNL}). The horizontal lines shown in the contour plots (\ref{contoura}) and (\ref{contourb}) represent the evolution trajectory of `$\Pi/\rho$' for given values of $\epsilon$, $w$, $\eta$, and a particular $\lambda$: the black line is for $\lambda=0.8$, the red line is for $\lambda=0.5$, and the purple line corresponds to $\lambda=0.2$. For $\lambda \ll 1$, results obtained in the linear ($n=0$) case becomes exactly applicable, i.e. acceptable solutions require, $\epsilon \ll 1$ and $\eta>1$. When $\lambda$ is large and approaches unity, the fluid, whose evolution begins in an out-of-equilibrium state, initially evolves even farther away from equilibrium before ultimately achieving the equilibrium state in the later phase, as illustrated in all three figures. In certain instances, the fluid may even exit the acceptable regime and enter the phantom era where, $\Pi/(\rho+p)\leq-1$ (as is the case of black curve in Fig. (\ref{PiNL})), before attaining equilibrium. While this kind of behavior implies the interesting scenario where the bulk viscous fluid crosses the phantom divide twice, such cases are plagued by causality violations. A similar case occurs when the values of $\eta$ and $\epsilon$ are fixed, and $\lambda$ is increased significantly (above unity). Here, it was observed that, at some finite energy density, viscous pressure diverges in the negative side, undergoes sign-switching, and then tend to zero from positive infinity, as seen from Fig. (\ref{PiNL}). As a result, this case also doesn't respect causality and should therefore be neglected from further analysis. The simplest way to ensure solutions in the acceptable regime is to constrain the fluid parameters with the limit; $\epsilon\ll1$, $\lambda<1/2$ and $\eta\geq1$. However, one must note that all these parameters are strongly correlated and one may also find acceptable solutions even outside these bounds.

\section{Dynamical system analysis of the coupled Einstein-Israel-Stewart system having causal relaxation time}\label{4}

	Evolution of bulk viscous fluid and the expansion of the Universe can be studied by solving the Israel-Stewart equation (\ref{diss}), which is coupled to the Einstein equations via the continuity relation (\ref{cons}). These two expressions together forms the coupled Einstein-Israel-Stewart system (EIS). By recasting these equations in terms of the dimensionless variables $\Omega_{\rho}=\rho/\Lambda_e$ and $\Omega_{\Pi}=\Pi/\Lambda_e$, we obtain the 2D non-linear autonomous system, 
\begin{align}
	\Omega_{\rho}^{\;\prime}&=f(\Omega_{\rho},\Omega_{\Pi})=-3\left[\left(1+w_0\right)\Omega_{\rho} + \Omega_{\Pi}\right]\label{ode3}\\
	\Omega_{\Pi}^{\;\prime}&=g(\Omega_{\rho},\Omega_{\Pi})=f\cdot\left\{\epsilon+\eta \left[ \frac{\Omega_{\Pi}\left(1+\lambda\Omega_{\Pi}\right)}{\Omega_{\rho}}\right]\right\}\label{ode4}
\end{align}
Here, overhead `\textit{prime}' denotes a derivative with respect to the number of e-folds $N$, defined as $N=\ln\left(a/a_i\right)$, with `$a_i$' being scale factor at beginning (which we set to one), and `$a$' is the scale factor at some later time. Clearly, the autonomous system offers (a) a fixed point\footnote{Points in $[\Omega_{\rho},\Omega_{\Pi}]$ phase plane where $\Omega^{\;\prime}_{\Pi}=\Omega^{\;\prime}_{\rho}=0$, are called the fixed/critical/equilibrium points of the autonomous system.}, `\textit{P}' where $\Omega_{\rho}=\Omega_{\Pi}=0$, and (b) a line of fixed points `\textit{L}' at which, $\Omega_{\Pi}=-(1+w_0)\Omega_{\rho}$. Note that the line of equilibrium points `\textit{L}' corresponds to regimes where the fluid is far from equilibrium where viscous pressure is significantly larger than the equilibrium pressure of the fluid. Therefore, if one expects the fluid to evolve from an out-of-equilibrium state in the early universe to a stable equilibrium state in the late phase, then \textit{L} must be a line of unstable equilibrium points, and \textit{P} must be a stable fixed point. In dynamical systems, (a) stable fixed points are those where all phase space trajectories converge, (b) unstable fixed points are those from which all trajectories diverge and (c) the points from which some trajectories converge, while some others diverge, are termed saddle. 

To determine the nature of fixed points, we construct the Jacobin matrix ($\mathcal{J}$) of the autonomous system and then determine its eigen values. For the system that we considering, $\mathcal{J}$ is defined as,
\begin{eqnarray}
	\mathcal{J}=\begin{bmatrix}
	\partial_{x} f & \partial_{y} f\\[10pt]
	\partial_{x} g  & \partial_{y} g 
\end{bmatrix} \quad \textbf{:} \quad x=\Omega_{\rho} \quad\& \quad y=\Omega_{\Pi}
\end{eqnarray}
Using Eqns. (\ref{ode3}) and (\ref{ode4}) we get,
\begin{align}
	\partial_x f&=-3(1+w_0)\quad &\textbf{:}&\quad \partial_y f= -3 \\
	\partial_x g&=f\partial_x Q+Q\partial_x f \quad &\textbf{:}&\quad \partial_y g =f\partial_y Q+Q\partial_y f
\end{align}
where we have defined, $g(x,y)=f(x,y)\cdot Q(x,y)$. To analyze the stability of the system near the fixed points, we only need to determine the nature of eigen values of $\mathcal{J}$ at those fixed points. Furthermore, we know that along the line of fixed points \textit{L}, and at the fixed point \textit{P}, we have $f=g=0$. This means, the jacobian matrix at the fixed points becomes,
\begin{figure}
	\centering
	\begin{subfigure}{0.495\columnwidth}
		\centering
		\includegraphics[width=\columnwidth]{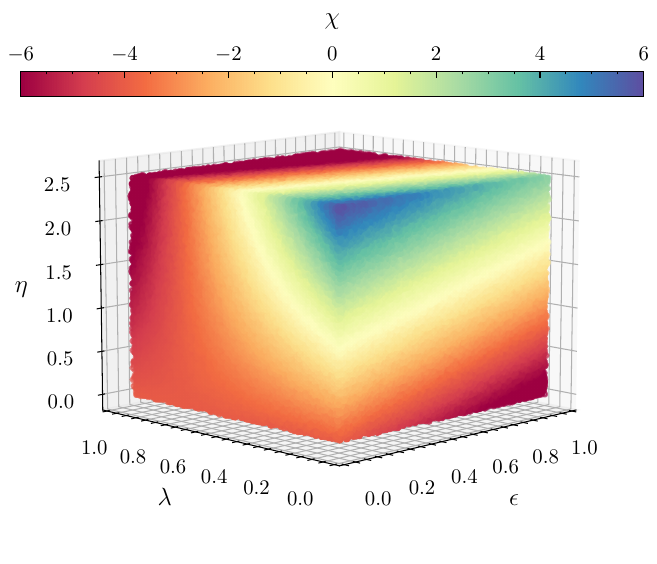}
		\caption{ }
		\label{eigenvalue}
	\end{subfigure}
	\begin{subfigure}{0.495\columnwidth}
		\centering
		\includegraphics[width=\columnwidth]{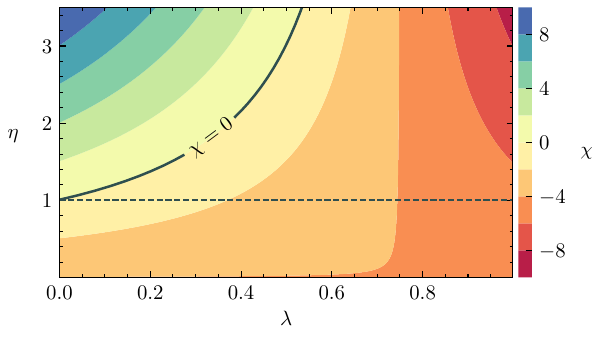}
		\caption{ }
		\label{eigencon}
	\end{subfigure}
	\caption{Plot of the eigenvalue $\chi$ that determines the stability of the system at the line of fixed points `\textit{L}'. Parameter space offering unstable fixed points are the ones with positive $\chi$ values, i.e. blue region. On the RHS, a contour plot of the eigenvalue $\chi$ for $\epsilon=0.01$ is also provided, i.e. Fig (\ref{eigencon}).}
\end{figure}

\begin{eqnarray}
	\mathcal{J}\;\bigg|_{\text{at fixed points}}=\begin{bmatrix}
		-3\left(1+w_0\right) & -3\\[10pt]
		-3\left(1+w_0\right)Q  & -3Q
	\end{bmatrix}
\end{eqnarray}
Analyzing the above matrix it is clear that the second row is basically $Q$ times the first row. Hence, the determinant of $\mathcal{J}$ at fixed points vanishes\footnote{This is due to the fact that the determinant of a matrix is equal to product of its eigen values} and as a result, one of the eigen values must be zero. Furthermore, since sum of the two eigen values must be the trace of the matrix $\mathcal{J}$, it is clear that the only eigen value of interest ($\chi$), that determines the stability of the system near fixed points, is simply the trace of $\mathcal{J}$ itself. That is,
\begin{equation}
	\chi=-3\left(1+w_0+Q\right)
\end{equation}
\noindent\textbf{(i)} At the fixed point `\textit{P}' at which $\Omega_{\Pi}=\Omega_{\rho}=0$ and $\Pi/\rho=-\kappa$, we have, $\chi=-3\left(1+w_0+\epsilon-\eta\kappa\right)$. Since $\eta\kappa\ll1$, the fixed point in the late phase becomes a stable attractor.

\noindent\textbf{(ii)} Along the line of fixed points `\textit{L}' we have the result, $\Omega_{\Pi}=-(1+w_0)\Omega_{\rho}$, and we get the eigen value,
\begin{equation}
	\chi=-3\left\{\epsilon+\left(1+w_0\right)\left[1-\eta\left(1-\lambda\left(1+w_0\right)\Omega_{\rho}\right)\right]\right\}
\end{equation}
Note that this expression can take both negative as well as positive signs depending on the value of the model parameters. Considering $w_0=1/3$ and $\Omega_{\rho}=1$, we plot this expression for analyzing the stability of fixed points as shown in Fig. (\ref{eigenvalue}) and (\ref{eigencon}). Accordingly, we see that the eigen value $\chi$ takes on positive values within the parameter space; $\lambda\ll1$, $\eta>1$ and $0<\epsilon\leq1$, and therefore, the points residing within this region predicts an unstable equilibrium state for the EIS system. While the region outside this parameter space has $\chi<0$, and therefore corresponds to an attractor. If one accepts the intuitive picture that, EIS system must be in an unstable equilibrium state when the dissipative fluid is far-from-equilibrium, then one must constrain the parameter space within the aforementioned limit, i.e. $\lambda\ll1$, $\eta>1$ and $0<\epsilon\leq1$. This ensures that trajectories in phase space diverge from \textit{L} and converge towards the stable fixed point \textit{P}. From the contour plot, Fig. (\ref{eigencon}), it is clear that for $\epsilon=0.01$, it is necessary that $\eta>1$ for having `\textit{L}' as an unstable fixed point.

\subsection{Slow-roll expansion driven by viscous cosmic fluid}

Considering the cosmic fluid in the early universe to be in a far from equilibrium state ($|\Pi| \approx \rho+p$), and the EIS system to be initialized very close to the unstable fixed point, we numerically solve  the coupled equations (\ref{ode3}) and (\ref{ode4}) by adopting a set of parameter values in the unstable region in the plot (\ref{eigenvalue}) where $\chi>0$. Then by using the obtained solution, we study the evolution characteristics of the Hubble parameter of the universe and its derivatives. To solve the coupled EIS system we consider\footnote{We choose this particular set of parameter values because they predict an accelerated expansion era that spans 50 efolds without requiring a fine-tuned value for the parameter $\mathfrak{w}$.}, $\epsilon=0.02$, $\eta=1.05$, $w_0=1/3$ along with the initial condition, $\Omega_{\Pi}=-\mathfrak{w}$ at $\Omega_{\rho}=1$, and investigate the cosmological dynamics in the early Universe. Note that this sample set of parameter values leads to $\chi=0.058$, and hence an unstable fixed point in the early universe.
\begin{figure}
	\centering
	\includegraphics[width=0.5\columnwidth]{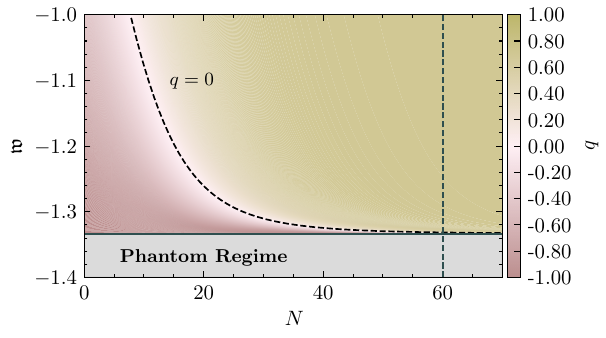}
	\caption{Contour plot of deceleration parameter of the universe with number of efolds `\textit{N}', for $\epsilon=0.02$, $\eta=1.05$, $w_0=1/3$ at different initial conditions $\mathfrak{w}$.}  
	\label{contourdec}
\end{figure}

\begin{figure}
	\centering
	\includegraphics[width=0.5\columnwidth]{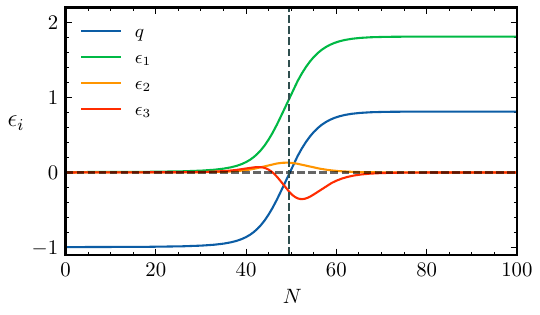}
	\caption{Evolution of Hubble slow-roll parameters for, $\epsilon=0.01$, $\eta=1.08$, $w_0=1/3$ and $\mathfrak{w}=1.33$.}
	\label{slowroll}
\end{figure}
In the far from equilibrium regime, the bulk viscosity associated with the effective cosmic fluid can be large enough to cause violation of the strong energy condition ($\rho+3p>0$) for a very small time, and hence cause an accelerated expansion of the universe. This is visualized from the plot of deceleration parameter `$q$' depicted in Fig. (\ref{contourdec}). There, we have expressed $q$ in terms of the number of efolds `$N=\ln(a)$' as, $q=-(1+d\ln(H)/dN)$. As the fluid relaxes from its initial out-of-equilibrium phase in the early universe to stable equilibrium in the late phase, the universe transitions from an accelerated to decelerated expansion phase. Furthermore, the exact transition point and the time interval of this accelerating phase depends on both, the value of model parameters, and the initial conditions applied on the model. Also, to produce a quasi-de-Sitter expansion driven by a viscous fluid having $c_s^2=w_0=1/3$, the initial conditions must be such that the value of the non-equilibrium parameter `$\mathfrak{w}$' is close to $4/3$. In addition to this, in fig. (\ref{contourdec}) we also see that in the case where $\epsilon=0.02$, $\eta=1.05$ and $w_0=1/3$, the value of non-equilibrium parameter $\mathfrak{w}$ must tend to $4/3$ for sustaining an accelerated expansion that spans 50 efolds or greater. Hence, as $\mathfrak{w}\approx 4/3$, (a) longer will be the time duration of the early accelerated expansion, and (b) more closer it resembles a de-Sitter phase.

The dynamics of the early universe, in particular the inflationary background, is better characterized using the Hubble slow-roll parameters, which forms a hierarchy, and is defined as the successive derivatives of the Hubble parameter with respect to the number of e-folds;
\begin{equation}
	\epsilon_{n+1} \equiv \frac{d \ln |\epsilon_{n}|}{dN},
\end{equation}
with the first one defined as, $\epsilon_{1} \equiv -\dot H/H^{2}$. These model independent parameters are significant as they directly determine the value of cosmological observables such as the scalar spectral tilt ($n_s$) and its running ($\alpha_s$) \cite{LiddleLyth,MukhanovBook}. The evolution of $\epsilon_{1}$, $\epsilon_{2}$, and $\epsilon_{3}$ with number of e-folds for acceptable values of free parameters mentioned earlier, is provided in Fig. \ref{slowroll}. Clearly, the parameter $\epsilon_{1}$ remains well below unity until $N\approx40$, confirming a background evolution that is deep in the accelerating regime. Then, its subsequent slow monotonic rise with the number of efolds indicates a gradual approach towards the end of inflation. The higher-order parameters $\epsilon_{2}$ and $\epsilon_{3}$ exhibit no particular oscillations/fluctuations around zero and therefore doesn't introduce unwanted instabilities. Hence, the scalar spectral tilt remains stable over the observable window, which can therefore yield predictions that are consistent with the Planck 2018 constraints \cite{aghanim2020planck}. Furthermore, the smallness of $\epsilon_{3}$ implies that the running of the spectral index ($\alpha_s$) is naturally negligible, which is also in agreement with current upper limits. The absence of abrupt jumps or sign-changing excursions in $\epsilon_{2}$ and $\epsilon_{3}$ also indicates a smooth evolution that is free from transient dynamical effects such as steps and turns. Overall, the graph illustrates a coherent and observationally viable slow-roll trajectory, equivalent to a stable single-field inflation and consistent with CMB constraints \cite{aghanim2020planck}. Finally, the evolution in Fig.~\ref{slowroll} clearly shows that $\epsilon_{1}$ rises smoothly from its inflationary value 
$\epsilon_1 \ll 1$, reaches the end-of-inflation condition $\epsilon_{1}=1$, and then approaches $\epsilon_{1}=2$. This final value corresponds to a radiation-dominated Universe with equation of state $\omega=1/3$. Thus, the model provides a smooth and natural transition from inflation to radiation domination, that is consistent with the expected thermal history of the early Universe.

\section{Results and Conclusion}

In this article, we derived an extended relation for the relaxation time of a barotropic Israel-Stewart fluid using the general nonlinear causality condition and analyzed its dynamical implications. One interesting outcome of this particular choice is that the nonlinear Israel-Stewart equation reduces to a first-order non-linear differential equation connecting the bulk viscous pressure and the energy density, which is valid in any homogeneous and isotropic spacetime. Furthermore, since the expression for the relaxation time of the fluid is directly derived from the causality constraint, this formulation can ensure causal evolution for the fluid, even when it is far from equilibrium. Hence, the dynamical equations governing the evolution of the fluid gets greatly simplified, without compromising causality.

For barotropic dissipative fluids in a spatially flat FLRW background, adopting this extended relaxation-time yields new class of exact analytical solutions in both the linear and nonlinear regimes. Notably, the effective equation of state resulting from these solutions predicts a generalized polytropic behavior. Dynamical systems analysis of the coupled Einstein–Israel–Stewart equations reveals that the solutions can admit acceptable cosmological solutions, with a line of unstable early time fixed points, and a stable late time attractor solution. This allows the coupled system to evolve away from the early far from equilibrium conditions and attain equilibrium in the late phase. Finally, we show that the present formulation can also be used to model a transient viscous-driven slow-roll expansion that smoothly exits into a radiation-dominated epoch, which is difficult to realize in standard inflationary scenarios.

Beyond the cosmological solutions explored here, the formalism developed in this work offers a versatile tool for probing dissipative phenomena across a broad range of relativistic systems. In astrophysics, the extended relaxation-time expression may refine models of bulk-viscous damping and thermalization in the interior of neutron-stars, proto-neutron-star evolution and post-merger dynamics \cite{Sawyer1989,PhysRevD.64.084003,Alford2010,Alford2018,Romatschke2010}, where significant departure from local equilibrium naturally may arise. In the regime of high-energy physics, the framework produced here may provide a consistent way to investigate nonlinear relaxation in the quark–gluon plasma, especially near the QCD critical region \cite{Romatschke2007,Heinz2013}. Furthermore, in early-universe cosmology, this formulation can enable a more realistic treatment of viscous or particle-producing phases during the QGP or reheating eras. Since equation for relaxation time is now fixed apriori by the causality constraint, the resulting dynamics becomes more robust. Hence, this formulation of general non-linear IS theory is particularly valuable for exploring far-from-equilibrium regimes where the choice of dissipative variables becomes ambiguous. 

\section*{Acknowledgments}

Vishnu A Pai thanks Cochin University of Science and Technology for providing Senior Research Fellowship. The authors acknowledge the assistance
of Grammarly \cite{grammarly} in improving the grammar and language usage in this article. After utilizing these tools, the authors thoroughly reviewed and edited the entire content, taking full responsibility for its accuracy and quality.


\begin{thebibliography}{47}%
	\makeatletter
	\providecommand \@ifxundefined [1]{%
		\@ifx{#1\undefined}
	}%
	\providecommand \@ifnum [1]{%
		\ifnum #1\expandafter \@firstoftwo
		\else \expandafter \@secondoftwo
		\fi
	}%
	\providecommand \@ifx [1]{%
		\ifx #1\expandafter \@firstoftwo
		\else \expandafter \@secondoftwo
		\fi
	}%
	\providecommand \natexlab [1]{#1}%
	\providecommand \enquote  [1]{``#1''}%
	\providecommand \bibnamefont  [1]{#1}%
	\providecommand \bibfnamefont [1]{#1}%
	\providecommand \citenamefont [1]{#1}%
	\providecommand \href@noop [0]{\@secondoftwo}%
	\providecommand \href [0]{\begingroup \@sanitize@url \@href}%
	\providecommand \@href[1]{\@@startlink{#1}\@@href}%
	\providecommand \@@href[1]{\endgroup#1\@@endlink}%
	\providecommand \@sanitize@url [0]{\catcode `\\12\catcode `\$12\catcode
		`\&12\catcode `\#12\catcode `\^12\catcode `\_12\catcode `\%12\relax}%
	\providecommand \@@startlink[1]{}%
	\providecommand \@@endlink[0]{}%
	\providecommand \url  [0]{\begingroup\@sanitize@url \@url }%
	\providecommand \@url [1]{\endgroup\@href {#1}{\urlprefix }}%
	\providecommand \urlprefix  [0]{URL }%
	\providecommand \Eprint [0]{\href }%
	\providecommand \doibase [0]{https://doi.org/}%
	\providecommand \selectlanguage [0]{\@gobble}%
	\providecommand \bibinfo  [0]{\@secondoftwo}%
	\providecommand \bibfield  [0]{\@secondoftwo}%
	\providecommand \translation [1]{[#1]}%
	\providecommand \BibitemOpen [0]{}%
	\providecommand \bibitemStop [0]{}%
	\providecommand \bibitemNoStop [0]{.\EOS\space}%
	\providecommand \EOS [0]{\spacefactor3000\relax}%
	\providecommand \BibitemShut  [1]{\csname bibitem#1\endcsname}%
	\let\auto@bib@innerbib\@empty
	\bibitem [{\citenamefont {Rocha}\ \emph {et~al.}(2024)\citenamefont {Rocha},
		\citenamefont {Wagner}, \citenamefont {Denicol}, \citenamefont {Noronha},\
		and\ \citenamefont {Rischke}}]{Rocha:2023ilf}%
	\BibitemOpen
	\bibfield  {author} {\bibinfo {author} {\bibfnamefont {G.~S.}\ \bibnamefont
			{Rocha}}, \bibinfo {author} {\bibfnamefont {D.}~\bibnamefont {Wagner}},
		\bibinfo {author} {\bibfnamefont {G.~S.}\ \bibnamefont {Denicol}}, \bibinfo
		{author} {\bibfnamefont {J.}~\bibnamefont {Noronha}},\ and\ \bibinfo {author}
		{\bibfnamefont {D.~H.}\ \bibnamefont {Rischke}},\ }\href
	{https://doi.org/10.3390/e26030189} {\bibfield  {journal} {\bibinfo
			{journal} {Entropy}\ }\textbf {\bibinfo {volume} {26}},\ \bibinfo {pages}
		{189} (\bibinfo {year} {2024})},\ \Eprint {https://arxiv.org/abs/2311.15063}
	{arXiv:2311.15063 [nucl-th]} \BibitemShut {NoStop}%
	\bibitem [{\citenamefont {Chabanov}\ \emph {et~al.}(2021)\citenamefont
		{Chabanov}, \citenamefont {Rezzolla},\ and\ \citenamefont
		{Rischke}}]{10.1093/mnras/stab1384}%
	\BibitemOpen
	\bibfield  {author} {\bibinfo {author} {\bibfnamefont {M.}~\bibnamefont
			{Chabanov}}, \bibinfo {author} {\bibfnamefont {L.}~\bibnamefont {Rezzolla}},\
		and\ \bibinfo {author} {\bibfnamefont {D.~H.}\ \bibnamefont {Rischke}},\
	}\href {https://doi.org/10.1093/mnras/stab1384} {\bibfield  {journal}
		{\bibinfo  {journal} {Monthly Notices of the Royal Astronomical Society}\
		}\textbf {\bibinfo {volume} {505}},\ \bibinfo {pages} {5910} (\bibinfo {year}
		{2021})},\ \Eprint
	{https://arxiv.org/abs/https://academic.oup.com/mnras/article-pdf/505/4/5910/38873579/stab1384.pdf}
	{https://academic.oup.com/mnras/article-pdf/505/4/5910/38873579/stab1384.pdf}
	\BibitemShut {NoStop}%
	\bibitem [{\citenamefont {Romatschke}\ and\ \citenamefont
		{Romatschke}(2019)}]{Romatschke:2017ejr}%
	\BibitemOpen
	\bibfield  {author} {\bibinfo {author} {\bibfnamefont {P.}~\bibnamefont
			{Romatschke}}\ and\ \bibinfo {author} {\bibfnamefont {U.}~\bibnamefont
			{Romatschke}},\ }\href {https://doi.org/10.1017/9781108651998} {\emph
		{\bibinfo {title} {{Relativistic Fluid Dynamics In and Out of
					Equilibrium}}}},\ Cambridge Monographs on Mathematical Physics\ (\bibinfo
	{publisher} {Cambridge University Press},\ \bibinfo {year} {2019})\ \Eprint
	{https://arxiv.org/abs/1712.05815} {arXiv:1712.05815 [nucl-th]} \BibitemShut
	{NoStop}%
	\bibitem [{\citenamefont {Maartens}(1995)}]{Maartens1995}%
	\BibitemOpen
	\bibfield  {author} {\bibinfo {author} {\bibfnamefont {R.}~\bibnamefont
			{Maartens}},\ }\href@noop {} {\bibfield  {journal} {\bibinfo  {journal}
			{Class. Quant. Grav.}\ }\textbf {\bibinfo {volume} {12}},\ \bibinfo {pages}
		{1455} (\bibinfo {year} {1995})}\BibitemShut {NoStop}%
	\bibitem [{\citenamefont {Zimdahl}(1996{\natexlab{a}})}]{Zimdahl1996}%
	\BibitemOpen
	\bibfield  {author} {\bibinfo {author} {\bibfnamefont {W.}~\bibnamefont
			{Zimdahl}},\ }\href@noop {} {\bibfield  {journal} {\bibinfo  {journal} {Phys.
				Rev. D}\ }\textbf {\bibinfo {volume} {53}},\ \bibinfo {pages} {5483}
		(\bibinfo {year} {1996}{\natexlab{a}})}\BibitemShut {NoStop}%
	\bibitem [{\citenamefont {Barrow}(1986)}]{Barrow1986}%
	\BibitemOpen
	\bibfield  {author} {\bibinfo {author} {\bibfnamefont {J.~D.}\ \bibnamefont
			{Barrow}},\ }\href@noop {} {\bibfield  {journal} {\bibinfo  {journal} {Phys.
				Lett. B}\ }\textbf {\bibinfo {volume} {180}},\ \bibinfo {pages} {335}
		(\bibinfo {year} {1986})}\BibitemShut {NoStop}%
	\bibitem [{\citenamefont {Lima}\ and\ \citenamefont
		{Germano}(1992)}]{Lima1996}%
	\BibitemOpen
	\bibfield  {author} {\bibinfo {author} {\bibfnamefont {J.~A.~S.}\
			\bibnamefont {Lima}}\ and\ \bibinfo {author} {\bibfnamefont {A.~S.~M.}\
			\bibnamefont {Germano}},\ }\href@noop {} {\bibfield  {journal} {\bibinfo
			{journal} {Phys. Lett. A}\ }\textbf {\bibinfo {volume} {170}},\ \bibinfo
		{pages} {373} (\bibinfo {year} {1992})}\BibitemShut {NoStop}%
	\bibitem [{\citenamefont
		{Zimdahl}(1996{\natexlab{b}})}]{10.1093/mnras/280.4.1239}%
	\BibitemOpen
	\bibfield  {author} {\bibinfo {author} {\bibfnamefont {W.}~\bibnamefont
			{Zimdahl}},\ }\href {https://doi.org/10.1093/mnras/280.4.1239} {\bibfield
		{journal} {\bibinfo  {journal} {Monthly Notices of the Royal Astronomical
				Society}\ }\textbf {\bibinfo {volume} {280}},\ \bibinfo {pages} {1239}
		(\bibinfo {year} {1996}{\natexlab{b}})},\ \Eprint
	{https://arxiv.org/abs/https://academic.oup.com/mnras/article-pdf/280/4/1239/18539963/280-4-1239.pdf}
	{https://academic.oup.com/mnras/article-pdf/280/4/1239/18539963/280-4-1239.pdf}
	\BibitemShut {NoStop}%
	\bibitem [{\citenamefont {Udey}\ and\ \citenamefont
		{Israel}(1982{\natexlab{a}})}]{10.1093/mnras/199.4.1137}%
	\BibitemOpen
	\bibfield  {author} {\bibinfo {author} {\bibfnamefont {N.}~\bibnamefont
			{Udey}}\ and\ \bibinfo {author} {\bibfnamefont {W.}~\bibnamefont {Israel}},\
	}\href {https://doi.org/10.1093/mnras/199.4.1137} {\bibfield  {journal}
		{\bibinfo  {journal} {Monthly Notices of the Royal Astronomical Society}\
		}\textbf {\bibinfo {volume} {199}},\ \bibinfo {pages} {1137} (\bibinfo {year}
		{1982}{\natexlab{a}})},\ \Eprint
	{https://arxiv.org/abs/https://academic.oup.com/mnras/article-pdf/199/4/1137/2882103/mnras199-1137.pdf}
	{https://academic.oup.com/mnras/article-pdf/199/4/1137/2882103/mnras199-1137.pdf}
	\BibitemShut {NoStop}%
	\bibitem [{\citenamefont {Giovannini}(1999)}]{Giovannini1999}%
	\BibitemOpen
	\bibfield  {author} {\bibinfo {author} {\bibfnamefont {M.}~\bibnamefont
			{Giovannini}},\ }\href@noop {} {\bibfield  {journal} {\bibinfo  {journal}
			{Phys. Rev. D}\ }\textbf {\bibinfo {volume} {60}},\ \bibinfo {pages} {123511}
		(\bibinfo {year} {1999})}\BibitemShut {NoStop}%
	\bibitem [{\citenamefont {Velten}\ and\ \citenamefont
		{Schwarz}(2012)}]{Velten2012}%
	\BibitemOpen
	\bibfield  {author} {\bibinfo {author} {\bibfnamefont {H.}~\bibnamefont
			{Velten}}\ and\ \bibinfo {author} {\bibfnamefont {D.~J.}\ \bibnamefont
			{Schwarz}},\ }\href@noop {} {\bibfield  {journal} {\bibinfo  {journal}
			{JCAP}\ }\textbf {\bibinfo {volume} {09}},\ \bibinfo {pages}
		{016}}\BibitemShut {NoStop}%
	\bibitem [{\citenamefont {N.~Cruz}\ and\ \citenamefont
		{Peña}(2017)}]{Cruz2015}%
	\BibitemOpen
	\bibfield  {author} {\bibinfo {author} {\bibfnamefont {S.~L.}\ \bibnamefont
			{N.~Cruz}}\ and\ \bibinfo {author} {\bibfnamefont {F.}~\bibnamefont
			{Peña}},\ }\href@noop {} {\bibfield  {journal} {\bibinfo  {journal} {Phys.
				Lett. B}\ }\textbf {\bibinfo {volume} {774}},\ \bibinfo {pages} {169}
		(\bibinfo {year} {2017})}\BibitemShut {NoStop}%
	\bibitem [{\citenamefont {Sasidharan}\ and\ \citenamefont
		{Mathew}(2015)}]{Sasidharan2015}%
	\BibitemOpen
	\bibfield  {author} {\bibinfo {author} {\bibfnamefont {A.}~\bibnamefont
			{Sasidharan}}\ and\ \bibinfo {author} {\bibfnamefont {T.~K.}\ \bibnamefont
			{Mathew}},\ }\href@noop {} {\bibfield  {journal} {\bibinfo  {journal} {Eur.
				Phys. J. C}\ }\textbf {\bibinfo {volume} {75}},\ \bibinfo {pages} {348}
		(\bibinfo {year} {2015})}\BibitemShut {NoStop}%
	\bibitem [{\citenamefont {Eckart}(1940)}]{Eckart1940}%
	\BibitemOpen
	\bibfield  {author} {\bibinfo {author} {\bibfnamefont {C.}~\bibnamefont
			{Eckart}},\ }\href@noop {} {\bibfield  {journal} {\bibinfo  {journal} {Phys.
				Rev.}\ }\textbf {\bibinfo {volume} {58}},\ \bibinfo {pages} {919} (\bibinfo
		{year} {1940})}\BibitemShut {NoStop}%
	\bibitem [{\citenamefont {Landau}\ and\ \citenamefont
		{Lifshitz}(1987)}]{Landau1987}%
	\BibitemOpen
	\bibfield  {author} {\bibinfo {author} {\bibfnamefont {L.~D.}\ \bibnamefont
			{Landau}}\ and\ \bibinfo {author} {\bibfnamefont {E.~M.}\ \bibnamefont
			{Lifshitz}},\ }\href@noop {} {\emph {\bibinfo {title} {Fluid Mechanics}}}\
	(\bibinfo  {publisher} {Pergamon Press},\ \bibinfo {year} {1987})\BibitemShut
	{NoStop}%
	\bibitem [{\citenamefont {Israel}\ and\ \citenamefont
		{Stewart}(1979)}]{IsraelStewart1979}%
	\BibitemOpen
	\bibfield  {author} {\bibinfo {author} {\bibfnamefont {W.}~\bibnamefont
			{Israel}}\ and\ \bibinfo {author} {\bibfnamefont {J.~M.}\ \bibnamefont
			{Stewart}},\ }\href@noop {} {\bibfield  {journal} {\bibinfo  {journal} {Ann.
				Phys.}\ }\textbf {\bibinfo {volume} {118}},\ \bibinfo {pages} {341} (\bibinfo
		{year} {1979})}\BibitemShut {NoStop}%
	\bibitem [{\citenamefont {Hiscock}\ and\ \citenamefont
		{Lindblom}(1985)}]{PhysRevD.31.725}%
	\BibitemOpen
	\bibfield  {author} {\bibinfo {author} {\bibfnamefont {W.~A.}\ \bibnamefont
			{Hiscock}}\ and\ \bibinfo {author} {\bibfnamefont {L.}~\bibnamefont
			{Lindblom}},\ }\href {https://doi.org/10.1103/PhysRevD.31.725} {\bibfield
		{journal} {\bibinfo  {journal} {Phys. Rev. D}\ }\textbf {\bibinfo {volume}
			{31}},\ \bibinfo {pages} {725} (\bibinfo {year} {1985})}\BibitemShut
	{NoStop}%
	\bibitem [{\citenamefont {Israel}(1976)}]{Israel1976}%
	\BibitemOpen
	\bibfield  {author} {\bibinfo {author} {\bibfnamefont {W.}~\bibnamefont
			{Israel}},\ }\href@noop {} {\bibfield  {journal} {\bibinfo  {journal} {Ann.
				Phys.}\ }\textbf {\bibinfo {volume} {100}},\ \bibinfo {pages} {310} (\bibinfo
		{year} {1976})}\BibitemShut {NoStop}%
	\bibitem [{\citenamefont {Bemfica}\ \emph {et~al.}(2019)\citenamefont
		{Bemfica}, \citenamefont {Disconzi},\ and\ \citenamefont
		{Noronha}}]{PhysRevLett.122.221602}%
	\BibitemOpen
	\bibfield  {author} {\bibinfo {author} {\bibfnamefont {F.~S.}\ \bibnamefont
			{Bemfica}}, \bibinfo {author} {\bibfnamefont {M.~M.}\ \bibnamefont
			{Disconzi}},\ and\ \bibinfo {author} {\bibfnamefont {J.}~\bibnamefont
			{Noronha}},\ }\href {https://doi.org/10.1103/PhysRevLett.122.221602}
	{\bibfield  {journal} {\bibinfo  {journal} {Phys. Rev. Lett.}\ }\textbf
		{\bibinfo {volume} {122}},\ \bibinfo {pages} {221602} (\bibinfo {year}
		{2019})}\BibitemShut {NoStop}%
	\bibitem [{\citenamefont {Weinberg}(1971)}]{Weinberg}%
	\BibitemOpen
	\bibfield  {author} {\bibinfo {author} {\bibfnamefont {S.}~\bibnamefont
			{Weinberg}},\ }\href {https://doi.org/10.1086/151073} {\bibfield  {journal}
		{\bibinfo  {journal} {Astrophys. J.}\ }\textbf {\bibinfo {volume} {168}},\
		\bibinfo {pages} {175} (\bibinfo {year} {1971})}\BibitemShut {NoStop}%
	\bibitem [{\citenamefont {Murphy}(1973)}]{PhysRevD.8.4231}%
	\BibitemOpen
	\bibfield  {author} {\bibinfo {author} {\bibfnamefont {G.~L.}\ \bibnamefont
			{Murphy}},\ }\href {https://doi.org/10.1103/PhysRevD.8.4231} {\bibfield
		{journal} {\bibinfo  {journal} {Phys. Rev. D}\ }\textbf {\bibinfo {volume}
			{8}},\ \bibinfo {pages} {4231} (\bibinfo {year} {1973})}\BibitemShut
	{NoStop}%
	\bibitem [{\citenamefont {Udey}\ and\ \citenamefont
		{Israel}(1982{\natexlab{b}})}]{udey}%
	\BibitemOpen
	\bibfield  {author} {\bibinfo {author} {\bibfnamefont {N.}~\bibnamefont
			{Udey}}\ and\ \bibinfo {author} {\bibfnamefont {W.}~\bibnamefont {Israel}},\
	}\href {https://doi.org/10.1093/mnras/199.4.1137} {\bibfield  {journal}
		{\bibinfo  {journal} {Monthly Notices of the Royal Astronomical Society}\
		}\textbf {\bibinfo {volume} {199}},\ \bibinfo {pages} {1137} (\bibinfo {year}
		{1982}{\natexlab{b}})},\ \Eprint
	{https://arxiv.org/abs/https://academic.oup.com/mnras/article-pdf/199/4/1137/2882103/mnras199-1137.pdf}
	{https://academic.oup.com/mnras/article-pdf/199/4/1137/2882103/mnras199-1137.pdf}
	\BibitemShut {NoStop}%
	\bibitem [{\citenamefont {Maartens}(1996)}]{Maartens1996}%
	\BibitemOpen
	\bibfield  {author} {\bibinfo {author} {\bibfnamefont {R.}~\bibnamefont
			{Maartens}},\ }\href@noop {} {\bibfield  {journal} {\bibinfo  {journal}
			{arXiv:astro-ph/9609119}\ } (\bibinfo {year} {1996})}\BibitemShut {NoStop}%
	\bibitem [{\citenamefont {Kremer}(2003)}]{Kremer2003}%
	\BibitemOpen
	\bibfield  {author} {\bibinfo {author} {\bibfnamefont {G.~M.}\ \bibnamefont
			{Kremer}},\ }\href@noop {} {\bibfield  {journal} {\bibinfo  {journal} {Phys.
				Rev. D}\ }\textbf {\bibinfo {volume} {68}},\ \bibinfo {pages} {123507}
		(\bibinfo {year} {2003})}\BibitemShut {NoStop}%
	\bibitem [{\citenamefont {Peralta-Ramos}\ and\ \citenamefont
		{Calzetta}(2012)}]{PeraltaRamos2012}%
	\BibitemOpen
	\bibfield  {author} {\bibinfo {author} {\bibfnamefont {J.}~\bibnamefont
			{Peralta-Ramos}}\ and\ \bibinfo {author} {\bibfnamefont {E.}~\bibnamefont
			{Calzetta}},\ }\href@noop {} {\bibfield  {journal} {\bibinfo  {journal}
			{Phys. Rev. D}\ }\textbf {\bibinfo {volume} {86}},\ \bibinfo {pages} {125024}
		(\bibinfo {year} {2012})}\BibitemShut {NoStop}%
	\bibitem [{\citenamefont {Barrow}(1990)}]{Barrow1990}%
	\BibitemOpen
	\bibfield  {author} {\bibinfo {author} {\bibfnamefont {J.~D.}\ \bibnamefont
			{Barrow}},\ }\href@noop {} {\bibfield  {journal} {\bibinfo  {journal} {Phys.
				Lett. B}\ }\textbf {\bibinfo {volume} {235}},\ \bibinfo {pages} {40}
		(\bibinfo {year} {1990})}\BibitemShut {NoStop}%
	\bibitem [{\citenamefont {Kamenshchik}\ \emph {et~al.}(2001)\citenamefont
		{Kamenshchik}, \citenamefont {Moschella},\ and\ \citenamefont
		{Pasquier}}]{KAMENSHCHIK2001265}%
	\BibitemOpen
	\bibfield  {author} {\bibinfo {author} {\bibfnamefont {A.}~\bibnamefont
			{Kamenshchik}}, \bibinfo {author} {\bibfnamefont {U.}~\bibnamefont
			{Moschella}},\ and\ \bibinfo {author} {\bibfnamefont {V.}~\bibnamefont
			{Pasquier}},\ }\href
	{https://doi.org/https://doi.org/10.1016/S0370-2693(01)00571-8} {\bibfield
		{journal} {\bibinfo  {journal} {Physics Letters B}\ }\textbf {\bibinfo
			{volume} {511}},\ \bibinfo {pages} {265} (\bibinfo {year}
		{2001})}\BibitemShut {NoStop}%
	\bibitem [{\citenamefont {CAMPO}\ and\ \citenamefont
		{VILLANUEVA}(2009)}]{doi:10.1142/S0218271809015795}%
	\BibitemOpen
	\bibfield  {author} {\bibinfo {author} {\bibfnamefont {S.~D.}\ \bibnamefont
			{CAMPO}}\ and\ \bibinfo {author} {\bibfnamefont {J.~R.}\ \bibnamefont
			{VILLANUEVA}},\ }\href {https://doi.org/10.1142/S0218271809015795} {\bibfield
		{journal} {\bibinfo  {journal} {International Journal of Modern Physics D}\
		}\textbf {\bibinfo {volume} {18}},\ \bibinfo {pages} {2007} (\bibinfo {year}
		{2009})},\ \Eprint
	{https://arxiv.org/abs/https://doi.org/10.1142/S0218271809015795}
	{https://doi.org/10.1142/S0218271809015795} \BibitemShut {NoStop}%
	\bibitem [{\citenamefont {Nojiri}\ \emph {et~al.}(2005)\citenamefont {Nojiri},
		\citenamefont {Odintsov},\ and\ \citenamefont
		{Tsujikawa}}]{PhysRevD.71.063004}%
	\BibitemOpen
	\bibfield  {author} {\bibinfo {author} {\bibfnamefont {S.}~\bibnamefont
			{Nojiri}}, \bibinfo {author} {\bibfnamefont {S.~D.}\ \bibnamefont
			{Odintsov}},\ and\ \bibinfo {author} {\bibfnamefont {S.}~\bibnamefont
			{Tsujikawa}},\ }\href {https://doi.org/10.1103/PhysRevD.71.063004} {\bibfield
		{journal} {\bibinfo  {journal} {Phys. Rev. D}\ }\textbf {\bibinfo {volume}
			{71}},\ \bibinfo {pages} {063004} (\bibinfo {year} {2005})}\BibitemShut
	{NoStop}%
	\bibitem [{\citenamefont {Chavanis}(2012)}]{Chavanis:2012uq}%
	\BibitemOpen
	\bibfield  {author} {\bibinfo {author} {\bibfnamefont {P.-H.}\ \bibnamefont
			{Chavanis}},\ }\href@noop {} {\  (\bibinfo {year} {2012})},\ \Eprint
	{https://arxiv.org/abs/1208.1185} {arXiv:1208.1185 [astro-ph.CO]}
	\BibitemShut {NoStop}%
	\bibitem [{\citenamefont {Dunsby}\ \emph {et~al.}(2024)\citenamefont {Dunsby},
		\citenamefont {Luongo}, \citenamefont {Muccino},\ and\ \citenamefont
		{Pillay}}]{DUNSBY2024101563}%
	\BibitemOpen
	\bibfield  {author} {\bibinfo {author} {\bibfnamefont {P.~K.}\ \bibnamefont
			{Dunsby}}, \bibinfo {author} {\bibfnamefont {O.}~\bibnamefont {Luongo}},
		\bibinfo {author} {\bibfnamefont {M.}~\bibnamefont {Muccino}},\ and\ \bibinfo
		{author} {\bibfnamefont {V.}~\bibnamefont {Pillay}},\ }\href
	{https://doi.org/https://doi.org/10.1016/j.dark.2024.101563} {\bibfield
		{journal} {\bibinfo  {journal} {Physics of the Dark Universe}\ }\textbf
		{\bibinfo {volume} {46}},\ \bibinfo {pages} {101563} (\bibinfo {year}
		{2024})}\BibitemShut {NoStop}%
	\bibitem [{\citenamefont {Chavanis}(2018)}]{Chavanis_2018}%
	\BibitemOpen
	\bibfield  {author} {\bibinfo {author} {\bibfnamefont {P.-H.}\ \bibnamefont
			{Chavanis}},\ }\href {https://doi.org/10.1088/1742-6596/1030/1/012009}
	{\bibfield  {journal} {\bibinfo  {journal} {Journal of Physics: Conference
				Series}\ }\textbf {\bibinfo {volume} {1030}},\ \bibinfo {pages} {012009}
		(\bibinfo {year} {2018})}\BibitemShut {NoStop}%
	\bibitem [{\citenamefont {Chimento}\ and\ \citenamefont
		{Jakubi}(1997)}]{Chimento:1997vy}%
	\BibitemOpen
	\bibfield  {author} {\bibinfo {author} {\bibfnamefont {L.~P.}\ \bibnamefont
			{Chimento}}\ and\ \bibinfo {author} {\bibfnamefont {A.~S.}\ \bibnamefont
			{Jakubi}},\ }\href {https://doi.org/10.1088/0264-9381/14/7/016} {\bibfield
		{journal} {\bibinfo  {journal} {Class. Quant. Grav.}\ }\textbf {\bibinfo
			{volume} {14}},\ \bibinfo {pages} {1811} (\bibinfo {year} {1997})},\ \Eprint
	{https://arxiv.org/abs/gr-qc/9703069} {arXiv:gr-qc/9703069} \BibitemShut
	{NoStop}%
	\bibitem [{\citenamefont {Di~Prisco}\ \emph {et~al.}(2000)\citenamefont
		{Di~Prisco}, \citenamefont {Herrera},\ and\ \citenamefont
		{Ib\'a\~nez}}]{PhysRevD.63.023501}%
	\BibitemOpen
	\bibfield  {author} {\bibinfo {author} {\bibfnamefont {A.}~\bibnamefont
			{Di~Prisco}}, \bibinfo {author} {\bibfnamefont {L.}~\bibnamefont {Herrera}},\
		and\ \bibinfo {author} {\bibfnamefont {J.}~\bibnamefont {Ib\'a\~nez}},\
	}\href {https://doi.org/10.1103/PhysRevD.63.023501} {\bibfield  {journal}
		{\bibinfo  {journal} {Phys. Rev. D}\ }\textbf {\bibinfo {volume} {63}},\
		\bibinfo {pages} {023501} (\bibinfo {year} {2000})}\BibitemShut {NoStop}%
	\bibitem [{\citenamefont {Cruz}\ \emph {et~al.}(2019)\citenamefont {Cruz},
		\citenamefont {Hern\'andez-Almada},\ and\ \citenamefont
		{Cornejo-P\'erez}}]{PhysRevD.100.083524}%
	\BibitemOpen
	\bibfield  {author} {\bibinfo {author} {\bibfnamefont {N.}~\bibnamefont
			{Cruz}}, \bibinfo {author} {\bibfnamefont {A.}~\bibnamefont
			{Hern\'andez-Almada}},\ and\ \bibinfo {author} {\bibfnamefont
			{O.}~\bibnamefont {Cornejo-P\'erez}},\ }\href
	{https://doi.org/10.1103/PhysRevD.100.083524} {\bibfield  {journal} {\bibinfo
			{journal} {Phys. Rev. D}\ }\textbf {\bibinfo {volume} {100}},\ \bibinfo
		{pages} {083524} (\bibinfo {year} {2019})}\BibitemShut {NoStop}%
	\bibitem [{\citenamefont {Gavassino}\ and\ \citenamefont
		{Noronha}(2024)}]{PhysRevD.109.096040}%
	\BibitemOpen
	\bibfield  {author} {\bibinfo {author} {\bibfnamefont {L.}~\bibnamefont
			{Gavassino}}\ and\ \bibinfo {author} {\bibfnamefont {J.}~\bibnamefont
			{Noronha}},\ }\href {https://doi.org/10.1103/PhysRevD.109.096040} {\bibfield
		{journal} {\bibinfo  {journal} {Phys. Rev. D}\ }\textbf {\bibinfo {volume}
			{109}},\ \bibinfo {pages} {096040} (\bibinfo {year} {2024})}\BibitemShut
	{NoStop}%
	\bibitem [{\citenamefont {Liddle}\ and\ \citenamefont
		{Lyth}(2000)}]{LiddleLyth}%
	\BibitemOpen
	\bibfield  {author} {\bibinfo {author} {\bibfnamefont {A.~R.}\ \bibnamefont
			{Liddle}}\ and\ \bibinfo {author} {\bibfnamefont {D.~H.}\ \bibnamefont
			{Lyth}},\ }\href@noop {} {\emph {\bibinfo {title} {Cosmological Inflation and
				Large-Scale Structure}}}\ (\bibinfo  {publisher} {Cambridge University
		Press},\ \bibinfo {year} {2000})\BibitemShut {NoStop}%
	\bibitem [{\citenamefont {Mukhanov}(2005)}]{MukhanovBook}%
	\BibitemOpen
	\bibfield  {author} {\bibinfo {author} {\bibfnamefont {V.}~\bibnamefont
			{Mukhanov}},\ }\href@noop {} {\emph {\bibinfo {title} {Physical Foundations
				of Cosmology}}}\ (\bibinfo  {publisher} {Cambridge University Press},\
	\bibinfo {year} {2005})\BibitemShut {NoStop}%
	\bibitem [{\citenamefont {Aghanim}\ \emph {et~al.}(2020)\citenamefont
		{Aghanim}, \citenamefont {Akrami}, \citenamefont {Ashdown}, \citenamefont
		{Aumont}, \citenamefont {Baccigalupi}, \citenamefont {Ballardini},
		\citenamefont {Banday}, \citenamefont {Barreiro}, \citenamefont {Bartolo},
		\citenamefont {Basak} \emph {et~al.}}]{aghanim2020planck}%
	\BibitemOpen
	\bibfield  {author} {\bibinfo {author} {\bibfnamefont {N.}~\bibnamefont
			{Aghanim}}, \bibinfo {author} {\bibfnamefont {Y.}~\bibnamefont {Akrami}},
		\bibinfo {author} {\bibfnamefont {M.}~\bibnamefont {Ashdown}}, \bibinfo
		{author} {\bibfnamefont {J.}~\bibnamefont {Aumont}}, \bibinfo {author}
		{\bibfnamefont {C.}~\bibnamefont {Baccigalupi}}, \bibinfo {author}
		{\bibfnamefont {M.}~\bibnamefont {Ballardini}}, \bibinfo {author}
		{\bibfnamefont {A.}~\bibnamefont {Banday}}, \bibinfo {author} {\bibfnamefont
			{R.}~\bibnamefont {Barreiro}}, \bibinfo {author} {\bibfnamefont
			{N.}~\bibnamefont {Bartolo}}, \bibinfo {author} {\bibfnamefont
			{S.}~\bibnamefont {Basak}}, \emph {et~al.},\ }\href@noop {} {\bibfield
		{journal} {\bibinfo  {journal} {Astronomy \& Astrophysics}\ }\textbf
		{\bibinfo {volume} {641}},\ \bibinfo {pages} {A6} (\bibinfo {year}
		{2020})}\BibitemShut {NoStop}%
	\bibitem [{\citenamefont {Sawyer}(1989)}]{Sawyer1989}%
	\BibitemOpen
	\bibfield  {author} {\bibinfo {author} {\bibfnamefont {R.~F.}\ \bibnamefont
			{Sawyer}},\ }\href {https://doi.org/10.1103/PhysRevD.39.3804} {\bibfield
		{journal} {\bibinfo  {journal} {Phys. Rev. D}\ }\textbf {\bibinfo {volume}
			{39}},\ \bibinfo {pages} {3804} (\bibinfo {year} {1989})}\BibitemShut
	{NoStop}%
	\bibitem [{\citenamefont {Jones}(2001)}]{PhysRevD.64.084003}%
	\BibitemOpen
	\bibfield  {author} {\bibinfo {author} {\bibfnamefont {P.~B.}\ \bibnamefont
			{Jones}},\ }\href {https://doi.org/10.1103/PhysRevD.64.084003} {\bibfield
		{journal} {\bibinfo  {journal} {Phys. Rev. D}\ }\textbf {\bibinfo {volume}
			{64}},\ \bibinfo {pages} {084003} (\bibinfo {year} {2001})}\BibitemShut
	{NoStop}%
	\bibitem [{\citenamefont {Alford}\ \emph {et~al.}(2012)\citenamefont {Alford},
		\citenamefont {Mahmoodifar},\ and\ \citenamefont {Schwenzer}}]{Alford2010}%
	\BibitemOpen
	\bibfield  {author} {\bibinfo {author} {\bibfnamefont {M.~G.}\ \bibnamefont
			{Alford}}, \bibinfo {author} {\bibfnamefont {S.}~\bibnamefont
			{Mahmoodifar}},\ and\ \bibinfo {author} {\bibfnamefont {K.}~\bibnamefont
			{Schwenzer}},\ }\href {https://doi.org/10.1103/PhysRevD.85.044051} {\bibfield
		{journal} {\bibinfo  {journal} {Phys. Rev. D}\ }\textbf {\bibinfo {volume}
			{85}},\ \bibinfo {pages} {044051} (\bibinfo {year} {2012})}\BibitemShut
	{NoStop}%
	\bibitem [{\citenamefont {Alford}\ and\ \citenamefont
		{Harris}(2018)}]{Alford2018}%
	\BibitemOpen
	\bibfield  {author} {\bibinfo {author} {\bibfnamefont {M.~G.}\ \bibnamefont
			{Alford}}\ and\ \bibinfo {author} {\bibfnamefont {S.~P.}\ \bibnamefont
			{Harris}},\ }\href {https://doi.org/10.1103/PhysRevC.98.065806} {\bibfield
		{journal} {\bibinfo  {journal} {Phys. Rev. C}\ }\textbf {\bibinfo {volume}
			{98}},\ \bibinfo {pages} {065806} (\bibinfo {year} {2018})}\BibitemShut
	{NoStop}%
	\bibitem [{\citenamefont {Romatschke}(2009)}]{Romatschke2010}%
	\BibitemOpen
	\bibfield  {author} {\bibinfo {author} {\bibfnamefont {P.}~\bibnamefont
			{Romatschke}},\ }\href {https://doi.org/10.1088/0264-9381/27/2/025006}
	{\bibfield  {journal} {\bibinfo  {journal} {Classical and Quantum Gravity}\
		}\textbf {\bibinfo {volume} {27}},\ \bibinfo {pages} {025006} (\bibinfo
		{year} {2009})}\BibitemShut {NoStop}%
	\bibitem [{\citenamefont {Romatschke}\ and\ \citenamefont
		{Romatschke}(2007)}]{Romatschke2007}%
	\BibitemOpen
	\bibfield  {author} {\bibinfo {author} {\bibfnamefont {P.}~\bibnamefont
			{Romatschke}}\ and\ \bibinfo {author} {\bibfnamefont {U.}~\bibnamefont
			{Romatschke}},\ }\href {https://doi.org/10.1103/PhysRevLett.99.172301}
	{\bibfield  {journal} {\bibinfo  {journal} {Phys. Rev. Lett.}\ }\textbf
		{\bibinfo {volume} {99}},\ \bibinfo {pages} {172301} (\bibinfo {year}
		{2007})}\BibitemShut {NoStop}%
	\bibitem [{\citenamefont {Heinz}\ and\ \citenamefont
		{Snellings}(2013)}]{Heinz2013}%
	\BibitemOpen
	\bibfield  {author} {\bibinfo {author} {\bibfnamefont {U.}~\bibnamefont
			{Heinz}}\ and\ \bibinfo {author} {\bibfnamefont {R.}~\bibnamefont
			{Snellings}},\ }\href {https://doi.org/10.1146/annurev-nucl-102212-170540}
	{\bibfield  {journal} {\bibinfo  {journal} {Ann. Rev. Nucl. Part. Sci.}\
		}\textbf {\bibinfo {volume} {63}},\ \bibinfo {pages} {123} (\bibinfo {year}
		{2013})},\ \Eprint {https://arxiv.org/abs/1301.2826} {arXiv:1301.2826
		[nucl-th]} \BibitemShut {NoStop}%
	\bibitem [{\citenamefont {{Grammarly Inc.}}(2024)}]{grammarly}%
	\BibitemOpen
	\bibfield  {author} {\bibinfo {author} {\bibnamefont {{Grammarly Inc.}}},\
	}\href@noop {} {\bibinfo {title} {Grammarly}},\ \bibinfo {howpublished}
	{\url{https://www.grammarly.com}} (\bibinfo {year} {2024}),\ \bibinfo {note}
	{aI-powered writing assistant}\BibitemShut {NoStop}%
\end{thebibliography}

%

\end{document}